\documentclass[]{SciPost}
\usepackage{booktabs}
\usepackage{upgreek}
\usepackage{multirow}
\usepackage{amssymb}
\usepackage[T1]{fontenc}
\usepackage{dsfont}
\usepackage{cleveref}
\usepackage{xspace}
\usepackage{url}
\usepackage{float}
\usepackage{graphicx}
\usepackage{appendix}
\graphicspath{ {./figures/} }

\begin{document}

\newcommand{\Pythia}{\textsc{Pythia}\xspace}
\newcommand{\mcplots}{\url{mcplots.cern.ch}\xspace}

\begin{center}{\Large \textbf{
String Fragmentation with a Time-Dependent Tension
}}\end{center}

\begin{center}
Nicholas Hunt-Smith\footnote{nicholashuntsmith@y7mail.com} and
Peter Skands\footnote{peter.skands@monash.edu}
\end{center}

\begin{center}
School of Physics and Astronomy, Monash University,\\
Wellington Road, Clayton, VIC-3800, Australia
\end{center}


\subsection*{Abstract}
{
Motivated by recent theoretical arguments that expanding strings can be regarded as having a temperature that is inversely proportional to the proper time, $\tau$, we investigate the consequences of adding a term $\propto 1/\tau$ to the string tension in the Lund string-hadronization model. The lattice value for the tension, $\kappa_0 \sim 0.18\,\mathrm{GeV}^2\sim 0.9\,\mathrm{GeV}/\mathrm{fm}$, is then interpreted as the late-time/equilibrium  limit. A generic prediction of this type of model is that early string breaks should be associated with higher strangeness (and baryon) fractions and higher fragmentation $\left<p_\perp\right>$ values. 
It should be possible to use archival $ee$ data sets to provide  model-independent constraints on this type of scenario, and we propose a few simple key measurements to do so.}


\section{Introduction}
\label{sec:intro}

Hadronization is an essential stage of high-energy particle collisions within QCD, describing how 
quarks and gluons  transform into observable hadrons. Although this process is intrinsically non-perturbative, a successful starting point for physical models has been the observation, e.g., in lattice calculations (see, e.g.,~\cite{Bali:1992ab}), that the potential between two static QCD charges (in an overall colour-singlet state) becomes asymptotically linear at long distances $r \gtrsim  1$~fm. This is the starting point for the Lund string fragmentation model~\cite{Andersson:1983ia,Sjostrand:1984ic,Andersson:1998tv}, which via its implementation in the \Pythia Monte Carlo event generator~\cite{Sjostrand:2006za,Sjostrand:2014zea} is among the most widely used  models today. 

Despite a track record of describing experimental measurements of particle production rates acceptably well in many contexts (see, e.g., the \mcplots web site~\cite{Karneyeu:2013aha}), it is worth keeping in mind that it remains a phenomenological model, whose underpinnings and simplifying assumptions can in principle be up for reevaluation in the light of new experimental results and/or new theoretical insights. 

One major simplifying assumption that has come under strong scrutiny in recent years is the fact that in the original formulation of the string-fragmentation model, each colour-singlet system is hadronized independently of any others. The mounting clear evidence of effects that appear to be collective in origin, including observations of flow-like effects~\cite{Khachatryan:2010gv,Velicanu:2011zz,Aad:2015gqa,Khachatryan:2015lva,Khachatryan:2016txc,Aaboud:2016yar,Aaboud:2017acw} and strangeness enhancements~\cite{Chatrchyan:2013qsa,ALICE:2017jyt,Acharya:2018orn,Aad:2019xek,Cui:2019jbt} in $pp$ collisions at the LHC, 
has spurred much activity in the theoretical community 
on how would-be independent string systems may fragment collectively in one way or another~\cite{Celik:1980td,Sjostrand:1993hi,Ferreiro:2003dw,Pierog:2013ria,Werner:2013tya,Werner:2014xoa,Bierlich:2014xba,Christiansen:2015yqa,Braun:2015eoa,Fischer:2016zzs,Bierlich:2016vgw,Bierlich:2017sxk,Bierlich:2017vhg,Citron:2018lsq,Bierlich:2018lbp,Duncan:2018gfk,Sjostrand:2018xcd,Bierlich:2019ixq,Nayak:2018xip,Bellm:2019wrh,Duncan:2019poz}. 
Common to these new modelling efforts is typically that they aim to account for the physical differences between fragmentation of a single string in isolation, such as in hadronic $Z$ decays, and the simultaneous hadronization of multiple such systems, such as in $pp$ or heavy-ion collisions, without making any 
direct modifications to the baseline single-string scenario. 

Given the renewed focus on the non-perturbative dynamics of hadronization, 
and the potential for new revealing measurements to be made at the LHC, 
it is worthwhile to reexamine also some of the foundations of the string model for the case of a single isolated string. There are of course very strong constraints from $e^+e^-\to\mathrm{hadrons}$, but to the 
extent that those constraints are mainly sensitive to various effective average quantities in the model, 
one may still ask questions such as whether a more dynamical picture could imply 
different fluctuations and/or different correlations without necessarily being excluded by the currently available constraints. Given the relative simplicity of the LEP data sets, 
such efforts may also be helpful in formulating new interesting measurements that could be made 
by groups with access to those data sets. 

In this work, we consider the implications of introducing an effective string tension which only asymptotically approaches the universal constant value observed on the lattice. There are at least three conceptual motivations for this:
\begin{itemize}
    \item The linear potential observed on the lattice and used as the starting point for the string model is, intrinsically, only a statement about the steady-state long-distance/late-time limit.
    It would not be invalidated by introducing departures from the strictly linear behaviour at early times, 
    short distances, and/or in non-equilibrium situations. 
    \item At short distances / early times, both lattice and perturbative QCD agree that there is a Coulomb component to the potential, which makes the potential well deeper than that of the simple linear asymptote. The asymptotic short-distance limit should be accounted for via perturbative $g\to q\bar{q}$ splittings in the parton shower, but an effective non-linearity may remain, associated with the transition region between the perturbative shower and the non-perturbative long-distance limit. Allowing for a larger effective string tension at early (proper) times may be a first (crude) way of mimicking the effect of a larger initial potential gradient.  
    We freely admit, however, that it is not obvious to us that the string description as such remains appropriate in the presence of such contributions. 
    \item It was recently highlighted~\cite{Berges:2017hne} that an expanding string may differ significantly from the steady-state situations studied on the lattice and in hadron spectroscopy. In particular, different regions of the string are necessarily entangled, implying the existence of a non-trivial entropy that increases with time. The string can therefore be interpreted as a thermal state, with a temperature that decreases with invariant (proper) time as $T \propto 1/\tau$. Without any attempt at further justification, we take this as an additional motivation to study a string with an effective tension $\kappa(\tau) \propto 1/\tau$. 
\end{itemize}

We note that the possibility of a \emph{fluctuating} string tension has been studied by other authors~\cite{Bialas:1999zg,Pirner:2018ccp}. The overall consequences are similar: broader hadron $p_\perp$ spectra and modified strangeness ratios, and we 
 expect that either type of model (as well as any of a similar ilk) could be constrained by the same set of experimental measurements. The possibility of a higher effective tension in the vicinity of heavy-quark endpoints has also been raised~\cite{Andersson:1981ce}. 
 Differences should show up in correlations (for instance with $N_\mathrm{ch}$), which will depend on the details of the proposed physics scenario, such as whether high string tensions can only occur at early times ($\tau$-dependent tension), at any time (fluctuating tensions), and/or for specific event topologies (heavy-flavour endpoints, gluon hairpin configurations, \ldots). 

This paper is organised as follows. In Sec.~\ref{sec:review}, we give an ultra-brief summary of the properties of the Lund string model that will be relevant for our study. In Sec.~\ref{sec:tauDep} we describe the modifications we make to the Lund model and define two example sets of toy-model parameters which are subsequently validated against a few salient LEP measurements in Sec.~\ref{sec:results}, in which we also propose a small set of new simple measurements that could be done on archival $ee$ data. Finally, in Sec.~\ref{sec:conclusions} we conclude and give an outlook to future work. 

\section{String Fragmentation Mini Review
\label{sec:review}}

The linear potential seen on the lattice at large distances is interpreted within the Lund model as the potential of a string, $V(r) = \kappa r$. Here, $r$ is the separation between the $q \bar{q}$ pair and $\kappa$ is ordinarily taken to be a constant string tension of around $0.18$ GeV$^2$ or $0.9$ GeV/fm. A stable meson consists of a $q \bar{q}$ pair which converts kinetic energy into potential energy stored in the string as the quarks move apart, before the quarks move back towards one another after exhausting all their kinetic energy and the process repeats. This oscillatory motion is called the \textit{Yoyo mode}. If there is enough energy present in the string, $q \bar{q}$ pairs can be created between the initial endpoint quarks, breaking the string into two distinct pieces. These breaks are modelled as a quantum mechanical tunnelling process, as originally devised by Schwinger in the context of $e^+e^-$ pair production in a strong electric field~\cite{Schwinger:1951}. This leads to a Gaussian distribution for the probability of producing a pair of quarks of mass $m_q$ and (oppositely oriented) transverse momentum, $p_\perp$:

\begin{equation}\label{eq:Schwinger}
    P(m_q^2, p_\perp^2) \propto \exp\left(-\frac{\pi m_q^2}{\kappa}\right) \exp\left(-\frac{\pi p_\perp^2}{\kappa}\right)~.
\end{equation}

String breaks will continue to occur until all remaining string pieces have sufficiently low invariant masses that they can be mapped onto hadronic states. This string fragmentation process is performed in hyperbolic coordinates, defined as~\cite[p.149]{Andersson:1998tv}:

\begin{equation}
    y=\frac{1}{2} \log{\frac{x_+}{x_-}}~,
\end{equation}

\begin{equation}\label{eq:Gamma}
    \Gamma=\kappa^2 x_+x_-~.
\end{equation}

Here, $x_{\pm}=t\pm x$ corresponds to light-cone space-time coordinates. The coordinate $y$ is then the rapidity, or the hyperbolic angle. Meanwhile, $\Gamma$ is related to the squared proper time of the vertex, $x_+x_-=t^2-x^2=\tau^2$. 

Individual string breaks are separated by spacelike intervals and are therefore causally disconnected, meaning that the string can be fragmented from the endpoints inwards rather than in a time-ordered fashion. Hadrons can then be split off iteratively from the endpoints in each step, according to the Lund symmetric fragmentation function~\cite{Andersson:1983ia}:
\begin{equation}
    f(z)=N\frac{1}{z} (1-z)^a\exp\left(\frac{-b(m_{h}^2 + p_{\perp h}^2)}{z}\right)~.
\end{equation}
This function defines the probability for a produced hadron of mass $m_h$ and transverse momentum $p_{\perp h}$ (relative to the string axis), to take a fraction $z$ of the remaining energy-momentum, and is defined so that the fragmentation process is  independent of which endpoint is chosen in each step. The parameters $a$ and $b$ are determined experimentally, while $N$ is a constant defined to make $f(z)$ integrate to unity. 

The probability distribution for $\Gamma$, and hence of $\tau = \sqrt{\Gamma}/\kappa$ via  eq.~(\ref{eq:Gamma}), is proportional to a modified area-law exponential~\cite{Andersson:1983ia},
\begin{equation}
    P(\Gamma) \propto \Gamma^a \exp(-b \Gamma)~. ~\label{eq:ProbGamma}
\end{equation}

\section{Time-Dependent Tension \label{sec:tauDep}}

Motivated by the finding~\cite{Berges:2017hne} that an expanding string may be described as having a temperature with an inverse dependence on proper time, we introduce a modified time-dependent string tension:

\begin{equation}
    \kappa(\tau) = \kappa_0 + \frac{A}{\mathrm{max}(\tau, \tau_0)}~,
\end{equation}
where $A$ sets the relative size (in GeV) of the $\tau$-dependent term,  $\kappa_0$ is the asymptotic string tension for late times which we take to be $\kappa_0 = 0.18\,\mathrm{GeV}^2 \sim 0.9\,\mathrm{GeV}/\mathrm{fm}$ 
in line with lattice results (e.g.,~\cite{Bali:1992ab}), $\tau$ is the proper time defined so that $\tau=0$ coincides with the production vertex of the hard (entangled) endpoint partons, and 
$\tau_0$ is a regularisation parameter that we introduce to ensure that the effective string tension has a maximum value at $\kappa_0 +\frac{A}{\tau_0}$ beyond which it is not allowed to increase any further. This prevents the model from generating unphysically large effects at very short distances ($\tau\to 0$ which we expect to be dominated by perturbative effects anyway). A natural choice for $\tau_0$ can be obtained by associating it with the inverse of the parton-shower cutoff in the model, which in PYTHIA by default is set at $0.5$ GeV (\verb|TimeShower:pTmin|), yielding a guess of
$\tau_0 = 2$ GeV$^{-1}$. 

An alternative way to constrain $\tau_0$ is by integrating the (time-dependent) potential energy stored in the string as a function of the quark-antiquark separation distance $R$, which for a general $\tau$-dependent tension can be expressed as:
\begin{equation}
    V(R) = 2 \int_{0}^{R/2}  \frac{\kappa(\tau)\,\tau\,\mathrm{d}\tau}{\sqrt{R^2/4-\tau^2}} ~, \label{eq:VeffR}
\end{equation}
and choosing $\tau_0$ such that the form of the Cornell potential~\cite{Eichten:1978tg} is approximately reproduced deeper into the Coulomb region. This is not a mandatory constraint, to the extent that the model is intended to represent physics (such as expansion) that may not be captured by the steady-state Cornell potential, but it furnishes an interesting option for a \emph{possible} constraint. Allowing the effective value of the strong coupling that normalises the Coulomb term in the Cornell potential to vary between its L\"uscher value, $V_\mathrm{Coulomb} = -\pi/(12R)$~\cite{Luscher:1980fr}, and the original Cornell value from charmonium $\sim -0.5/R$~\cite{Eichten:1978tg}, we find the following rough range for ``Coulomb-inspired''  $\tau_0$ values:
\begin{equation} \label{eq:tau0}
    \frac{1}{\pi^2 A} \lesssim \tau_0 \lesssim \frac{2}{\pi^2 A}~.
\end{equation}

As a final constraint, we require that the average value of $\kappa$ in string breaks in our toy scenarios should be the same as the effective $\kappa$ value used in the baseline (Monash 2013) tune of PYTHIA~\cite{Skands:2014pea}. This will ensure that average hadron-level observables will, at least to a first approximation, 
remain the same as in the Monash tune.

\begin{figure}[t]
    \centering
    \includegraphics[width=0.67\textwidth]{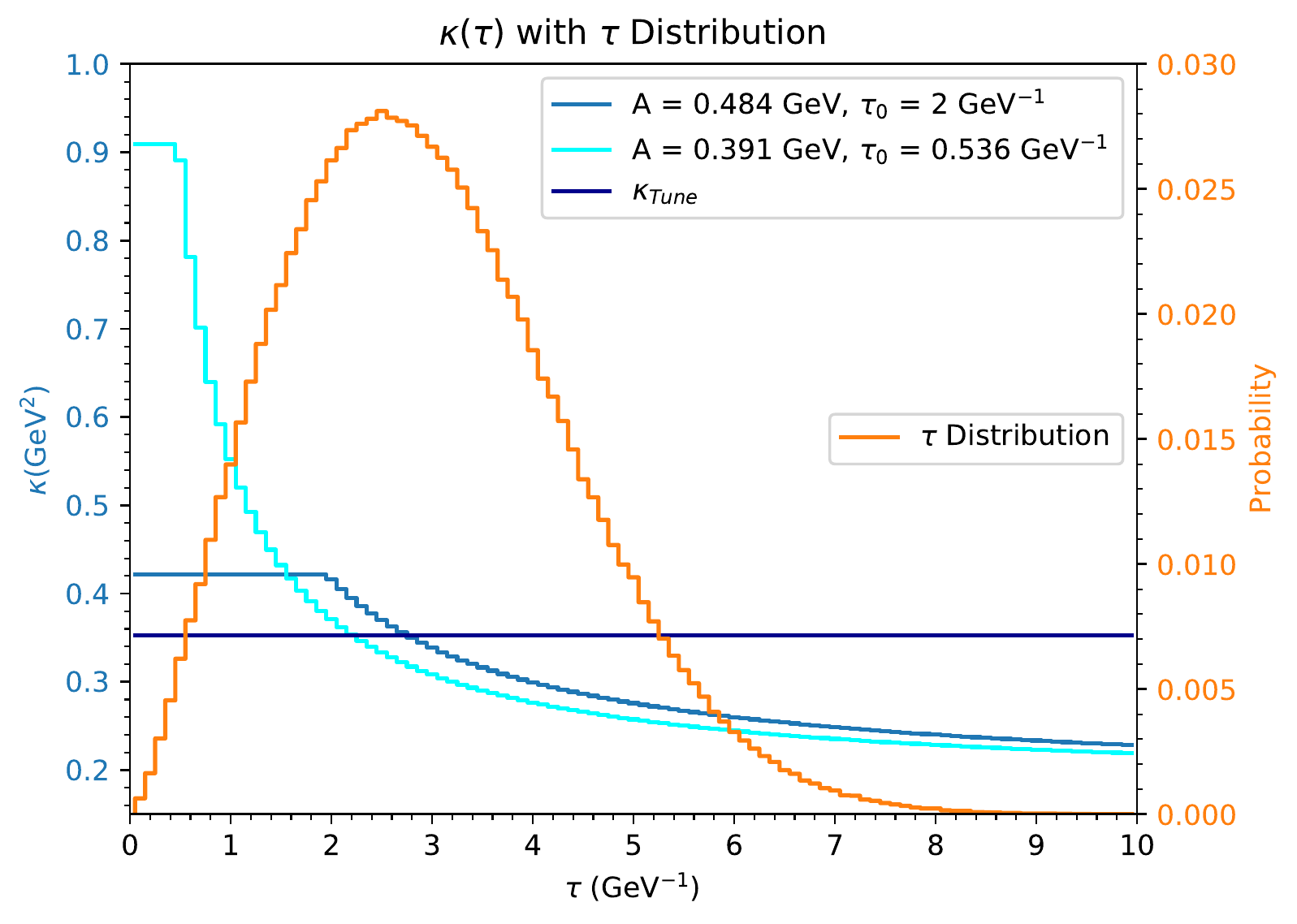}
    \caption{Modified string tension $\kappa(\tau)$ compared to constant tension $\kappa_{tune}$ with $\tau$ distribution for different definitions of $\tau_0$ and $A$.}
    \label{fig:kappaTau}
\end{figure}
The $\tau$ distribution for string breaks in $e^+e^-\to\mathrm{hadrons}$ is illustrated by the orange histogram in  fig.~\ref{fig:kappaTau}, for the reference tune. 
The corresponding constant value $\kappa_\mathrm{tune} = 0.353\,\mathrm{GeV}^2 \sim 1.8\,\mathrm{GeV}/\mathrm{fm}$ (obtained from the $p_\perp$ broadening of the reference tune, see below) 
is indicated by the dark blue horizontal line.
The two lighter-blue histograms illustrate two variants of our model constrained to keep  $\left<\kappa(\tau)\right> = \kappa_\mathrm{tune}$\footnote{Calculated using the same $\tau$ distribution as for the reference model. The actual $\tau$ distribution will be slightly broader in the $\tau$-dependent scenarios, but this has a small effect on the average value.}:
\begin{itemize}
    \item $A = 0.484$ GeV, $\tau_0 = 2$ GeV$^{-1}$,
    \item $A = 0.391$ GeV, $\tau_0 = 0.536$ GeV$^{-1}$,
\end{itemize}
where the former corresponds to letting the shower cutoff define $\tau_0$, with maximal tension $\kappa_\mathrm{max} = \kappa_0 + A/\tau_0 \sim 2.3 \kappa_0 \sim 1.2 \kappa_\mathrm{tune}$ (i.e., only slightly higher than the average value in the reference tune)
while the latter represents an example of using the  ``Coulomb-inspired'' constraint\footnote{Note that we allowed 
a slight violation of the bound expressed by eq.(\ref{eq:tau0}) to get a nice integer number for the ratio $\kappa_\mathrm{max}/\kappa_0$.}, 
chosen such that the maximal tension is substantially larger than in the reference tune, $\kappa_\mathrm{max} = 5 \kappa_0 \sim 2.5 \kappa_\mathrm{tune}$, with the tradeoff that it is then only active for a relatively short time. 

The $\tau$ distribution in fig.~\ref{fig:kappaTau} is determined by the parameters $a=0.68$ and $b=0.98$ of the reference tune via eq.~(\ref{eq:ProbGamma}). Reducing the $a$ parameter would in principle allow one to probe even earlier $\tau$ (and hence possibly larger variations in the effective string tension); but this would have to be accompanied by a similar reduction in $b$ to maintain the same average charged multiplicity. In practice, we did not find that variations with e.g.\ $a=0.23$ (a third of the reference value) and $b=0.49$ (half of the reference value) produced substantial changes in the distributions we explore in this work. Thus we show only results obtained with the reference values for $a$ and $b$ in what follows. 

\begin{figure}[tp]
\centering\includegraphics*[width=0.99\textwidth]{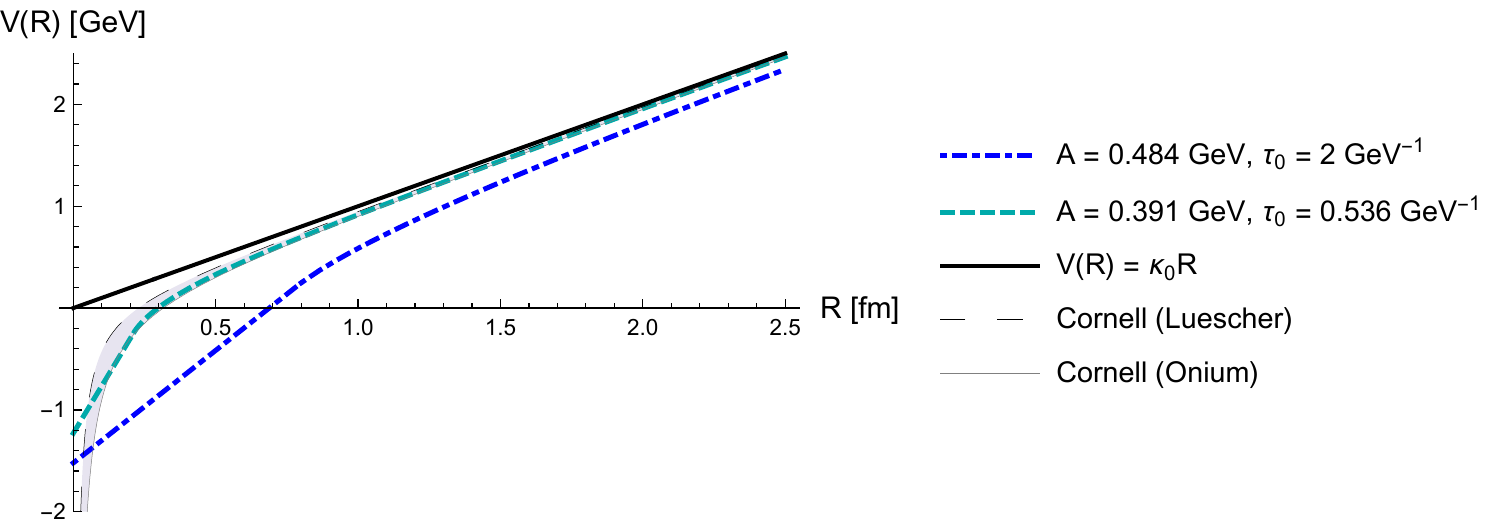}
\caption{The effective quark-antiquark potential, $V(R)$. The two example cases of $\tau$-dependence parameters used in our study (shown with dot-dashed blue and dashed cyan lines), compared with the standard linear part (solid black), and two variants of the Cornell potential (shaded gray band).}
\label{fig:cornell}
\end{figure}
In fig.~\ref{fig:cornell}, we compare the effective potentials $V(R)$ for these two toy scenarios (dot-dashed and dashed respectively) to the Cornell potential (grey shaded band), in all cases choosing the constant $V_0$ term so that the same behaviour is obtained for asymptotically long distances. It is clear that only the Coulomb-inspired (dashed cyan) model resembles the Cornell potential down to fairly low $R$ (beyond which, presumably, the physics is anyway fully perturbative). Since our model was not originally conceived to describe Coulomb effects, we emphasise that we do not regard either of these scenarios as more or less well motivated than the other. They simply represent a variation between allowing a slightly increased tension for a relatively long time (the shower-cutoff-motivated scenario) versus allowing a highly increased tension for a shorter time (the Coulomb-inspired scenario). Other combinations of $A$ and $\tau_0$ that result in the same $\left<\kappa(\tau)\right>$ are collected in tab.~\ref{tab:Avstau0} in the appendix. 
Whichever values of $A$, $\tau_0$, and $\kappa_0$ are used, 
a modified string tension will result in a modified $\kappa$ in eq.~\eqref{eq:Schwinger} and eq.~\eqref{eq:Gamma}. 

The first factor in eq.~\eqref{eq:Schwinger} relates to the yield ratio of strange quarks to up/down quarks:

\begin{equation}\label{eq:ProbSoUD}
    P(s:u/d) = \frac{P(m_s^2)}{P(m_{u/d}^2)} = \exp\left(-\frac{(m_s^2-m_{u/d}^2)}{\kappa}\right)~,
\end{equation}
where $m_s$ is the mass of the strange quark and $m_{u/d}$ is the mass of the up/down quarks. Since the values of these masses are highly scheme dependent and appear in an exponent here, the above formula is normally not useful for practical applications; instead, the approach in PYTHIA is to parameterise the strangeness suppression parameter, $P(s:u/d)$, directly and fit that to experimental data (see, e.g.,  \cite{Skands:2014pea}), resulting in a strangeness ratio of around 0.2-0.3 for a very wide range of CM energies~\cite{Skands:2014pea,Park:2014zra}. Under our modified string tension, at early times the strangeness ratio will be increased, and at later times it will be decreased. Denoting the average strangeness suppression by $    \left<P(s:u/d)\right> $, 
the suppression factor for a string break that occurs at (proper) time $\tau$ is:
\begin{equation}
   P(s:u/d)(\tau) =\left<P(s:u/d)\right>^{(\left<\kappa\right>/\kappa(\tau))}~.\label{eq:ProbStoUDeff}
 \end{equation}
 
For baryons, PYTHIA's modelling is based on string breaks involving diquarks, as described e.g.\ in~\cite[Section~12.1.3]{Sjostrand:2006za}. Analogously to what we do for strangeness, 
we introduce modifications to the diquark production parameters, following the method detailed in \cite{Bierlich:2014xba}.
Specifically, we assume that the suppression factor for strange diquarks relative to non-strange ones scales in the same way as in eq.~\eqref{eq:ProbStoUDeff}, and that this also applies to the rate of spin-1 diquarks relative to spin-0 ones. That then leaves only the scaling of the overall suppression factor for producing diquarks relative to quarks to be determined. This is complicated somewhat due to the underlying so-called ``popcorn'' mechanism of diquark production. According to the popcorn picture, there is first a $q\bar{q}$ fluctuation which leaves the string intact, with a probability we denote $\beta$, followed by a second $q\bar{q}$ fluctuation which tunnels out to break the string, with a probability $\gamma$. The overall diquark-to-quark suppression factor is then $\xi = \alpha \beta \gamma$, where $\alpha$ can be expressed in terms of the other suppression factors mentioned above:
\begin{equation} 
    \alpha = \frac{1 + 2x\rho + 9y + 6x\rho y + 3y x^2 \rho^2}{2+\rho}~,
\end{equation}
with $\rho$ and $x\rho$ the strangeness suppression factors for quarks and diquarks respectively, and $y$ the suppression factor for spin-1 diquarks relative to spin-0 ones. 
Since $\gamma$ is a tunnelling probability, we assume that it scales in the same way as the other suppression factors, i.e.\ as $\tilde{\gamma} = \left(\gamma\right)^{\left<\kappa\right>/\kappa(\tau)}$, while $\beta$ is the constant popcorn production probability. The modified diquark suppression factor ($\tilde{\xi}$) is therefore given by:
\begin{equation}
    \tilde{\xi} = \tilde{\alpha} \beta \tilde{\gamma} = \tilde{\alpha} \beta \left(\frac{\xi}{\alpha\beta}\right)^{\left<\kappa\right>/\kappa(\tau)}~.
\end{equation}
 
The second factor in eq.~\eqref{eq:Schwinger} relates to the width of the $p_\perp$ spectrum:
\begin{equation}
    \sigma^2 = \left<p_\perp^2\right> = \frac{\pi}{\kappa} \int_{0}^{\infty} p_\perp^2 \exp \left(\frac{-\pi  p_\perp^2}{\kappa}\right) d p_\perp^2~ = \frac{\kappa}{\pi}~.
    \label{eq:sigmaKappa}
\end{equation}
Inserting the $\sigma$ value of the Monash tune, \verb|StringPT:sigma = 0.335| (GeV), we get 
\begin{equation}
    \kappa_\mathrm{tune} = 0.353\,\mathrm{GeV}^2~\sim~1.8\,\mathrm{GeV}/\mathrm{fm},
\end{equation}
which we used to constrain the average tension, $\left<\kappa(\tau)\right>$, above. For late times, the lattice value  $\kappa_0 = 0.18\,\mathrm{GeV}^2 \sim 0.9\,\mathrm{GeV}/\mathrm{fm}$ corresponds to a $p_\perp$  broadening of $\sigma_0 = 0.24\,\mathrm{GeV}$, while the largest tension reached for $\tau\le\tau_0$ by the Coulomb-inspired  scenario, $\kappa_\mathrm{max} \sim 0.9 \,\mathrm{GeV}^2 \sim 4.5 \,\mathrm{GeV}/\mathrm{fm}$, 
corresponds to  $\sigma_\mathrm{max} = 0.54\,\mathrm{GeV}$. 

As with the strangeness ratio, the average $p_\perp$ will be unchanged in the scenarios we study here, but the $p_\perp$ given to individual hadrons will depend on the $\tau$ at which their corresponding string breaks occurred. This will alter the $p_\perp$ correlations among hadrons which share a particularly early (or late) string break, and it will also alter the $p_\perp$-strangeness correlations. 

In addition, the definition of the hyperbolic coordinate $\Gamma$ will change from eq.~\eqref{eq:Gamma} to:
\begin{equation}\label{eq:newGamma}
    \Gamma=\left(\kappa_0 + \frac{A}{\mathrm{max}(\tau, \tau_0)}\right)^2 \tau^2 ~,
\end{equation}
which maintains Lorentz invariance since $\tau$ is the proper time. 

In this work, we neglect possible modifications to the longitudinal aspect of the fragmentation. In principle, the $\tau$-dependent tension will lead to a different rate of energy loss $dE/dt$ for the endpoint quarks, by analogy with eq.~(\ref{eq:VeffR}), and this may in turn affect the longitudinal momentum spectrum of produced hadrons. However, since our model does still maintain the property of self-similarity (and hence left-right symmetry) for each (fixed) value of $\tau$ and since $f(z)$ does not depend explicitly on the string tension, we believe modifications to the longitudinal properties of the fragmentation should be small. Therefore, while we think it would be interesting to follow up on this question in a future study (especially if experimental measurements should support the predictions we make for the strangeness, baryon, and $p_\perp$ aspects), 
for now we let PYTHIA select $z$ and $\Gamma$ values using the $a$ and $b$ parameters of the reference tune without modifications. We then compute $\tau$ and $\kappa(\tau)$ using eq.~\eqref{eq:newGamma}. Finally, we modify the strangeness ratio and $p_\perp$ using eqs.~\eqref{eq:ProbStoUDeff} and \eqref{eq:sigmaKappa}.

At the technical level, we make use of the \verb|UserHook| functionality within PYTHIA, which allows to intervene at various stages of the hadronization process and modify the properties of each string break on an individual basis. This is important since each string break will have its own  $\tau$, so a blanket value cannot be specified for the strangeness ratio and $p_\perp$. Our modifications were introduced using the \verb|doChangeFragPar| method, which reinitialises the PYTHIA settings with our adjusted values. In this case, the relevant settings were \verb|StringFlav:probStoUD|, \verb|StringFlav:probQQtoQ|, \verb|StringFlav:probSQtoQQ|, \verb|StringFlav:probQQ1toQQ0| and 
\verb|StringPT:sigma|.

\section{Results \label{sec:results}}

The  results presented in this section were obtained by generating 4 million $e e \to \mathrm{hadrons}$ collisions at $\sqrt{s} = 91\,\mathrm{GeV}$, with standard LEP settings (ISR switched off, and particles with lifetimes $c\tau > 100\,\mathrm{mm}$ treated as stable). We use PYTHIA 8.243, augmented by an implementation of our model via the \text{UserHooks} framework, and compare the two example scenarios for $\tau_0$ discussed above to the baseline PYTHIA modelling. For reference, the toy scenarios we include are:
\begin{itemize}
    \item Toy Scenario 1: $A = 0.484\,\mathrm{GeV},~\tau_0 = 2\,\mathrm{GeV}^{-1},~\kappa_0=0.18\,\mathrm{Gev}^2$, with  $\tau_0$  set to the inverse of the parton shower cutoff and $\kappa_0$ set to its lattice value. 
    \item Toy Scenario 2: $A = 0.391\,\mathrm{GeV},\tau_0 = 0.536\,\mathrm{GeV}^{-1},~\kappa_0=0.18\,\mathrm{GeV}^2$, roughly motivated by producing a Coulomb-like deviation from the linear term in the potential. 
\end{itemize}
All parameters not explicitly mentioned are fixed to their Monash 2013 values~\cite{Skands:2014pea}, which is the default parameter set for all PYTHIA 8.2 versions. 

\begin{figure}[tp]
\centering
\hspace*{-2mm}\begin{tabular}{cc}
\includegraphics[width=0.48\textwidth]{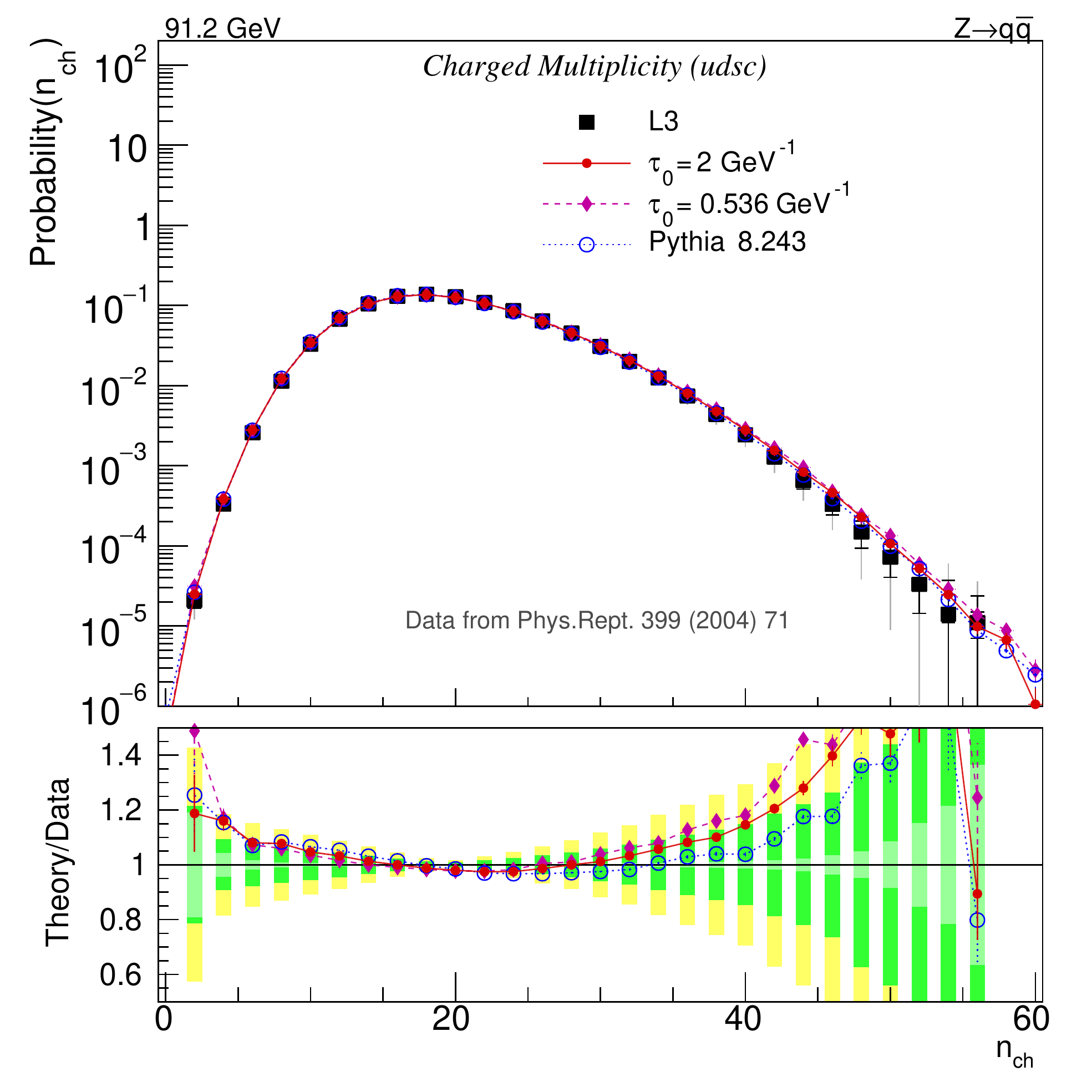} & 
\includegraphics[width=0.48\textwidth]{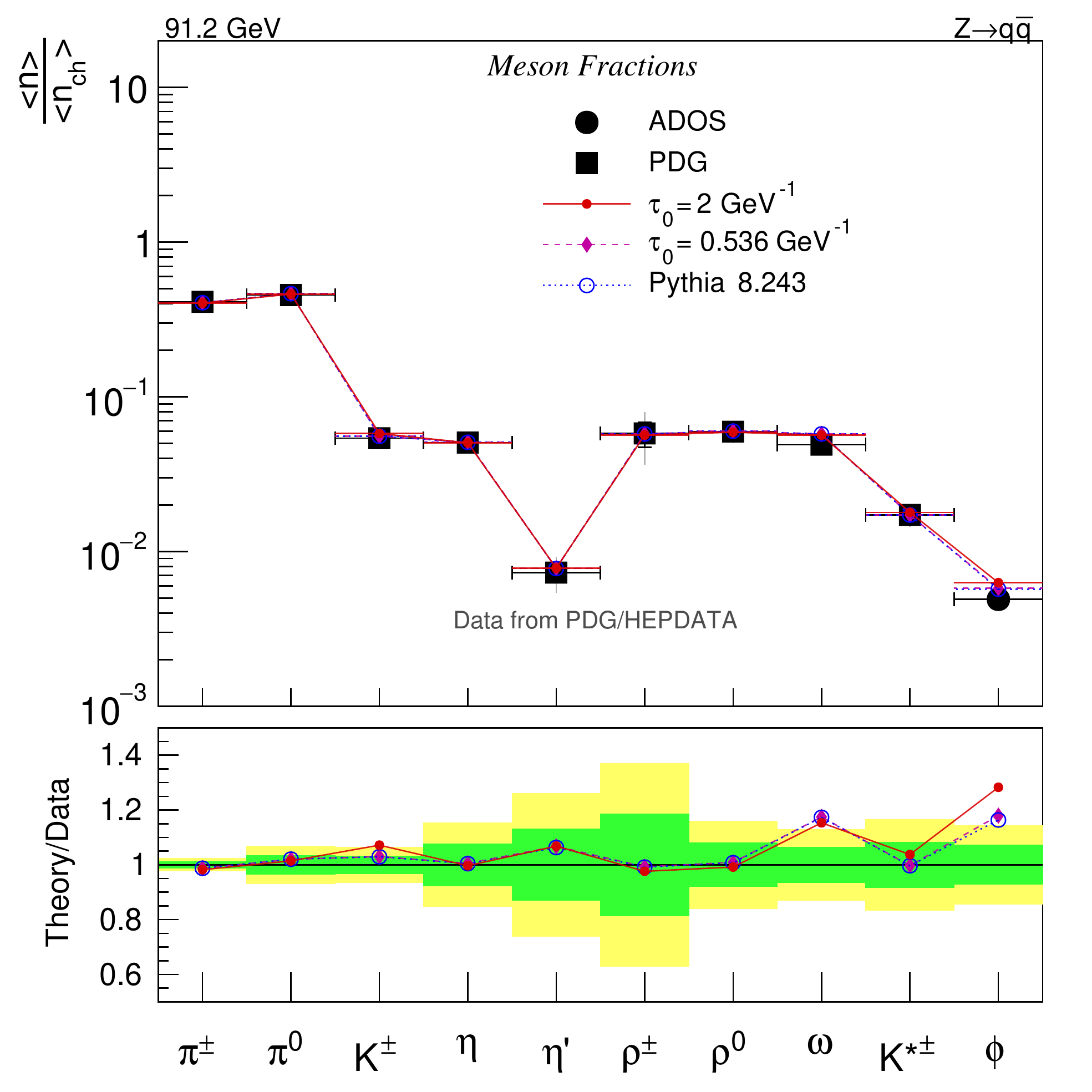} \\[2mm]
\includegraphics[width=0.48\textwidth]{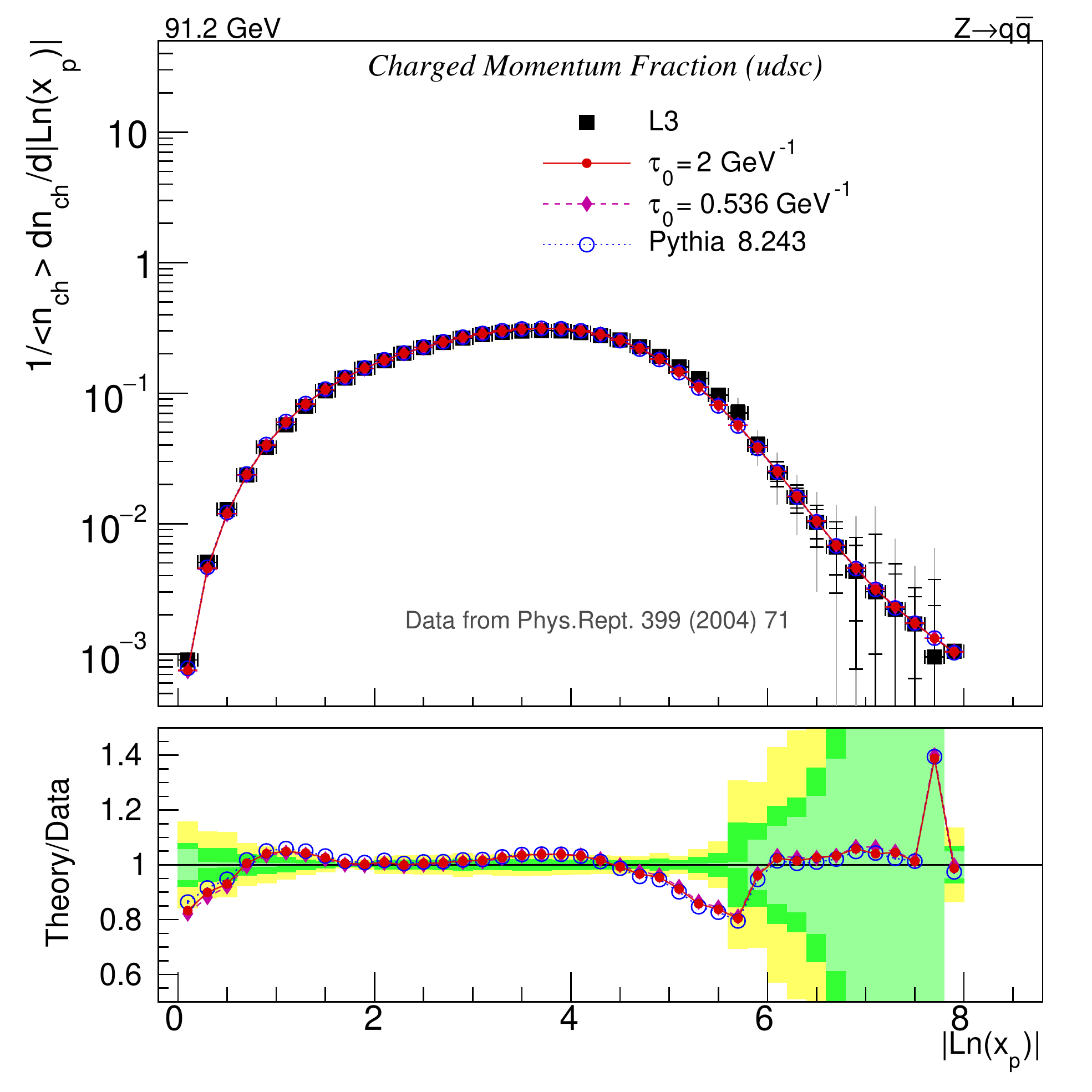} & 
\includegraphics[width=0.48\textwidth]{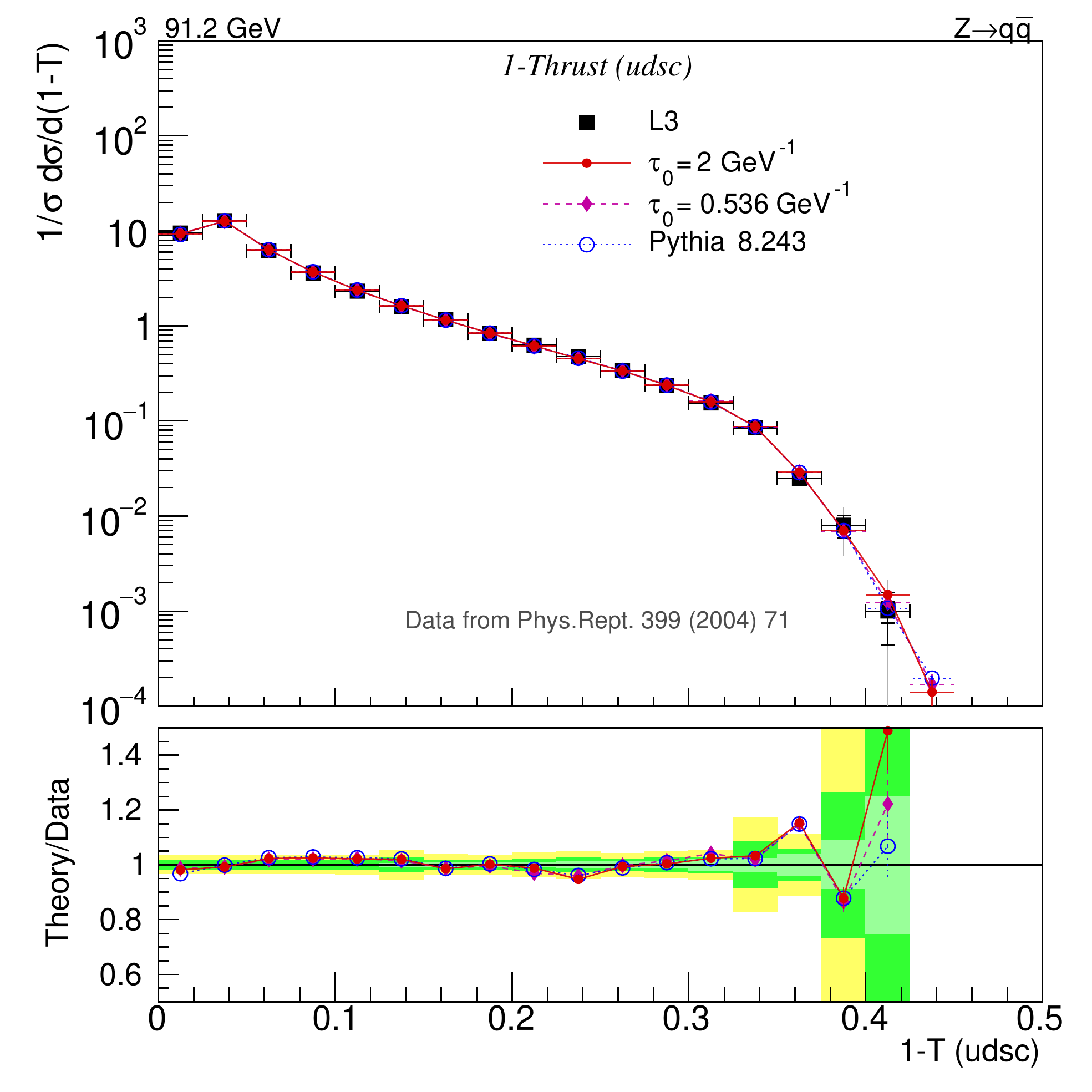} 
\end{tabular}
\caption{Overall event properties in hadronic $Z$ decays, specifically the charged multiplicity (top left), meson fractions (top right), charged momentum spectrum (bottom left), and Thrust distribution (bottom right). Data from~\cite{Achard:2004sv,Tanabashi:2018oca}. 
\label{fig:monashRef}}
\end{figure}

Firstly,  since our example scenarios are both constrained to give roughly the same average value of $\kappa$ as for the Monash tune, we expect overall event properties, such as average multiplicities and average strangeness fractions, to remain roughly the same.  
This is validated in fig.~\ref{fig:monashRef}, where we show the charged multiplicity distribution (top left), the production rates of various meson types (top right), the charged $x_p = 2 |p|/E_\mathrm{CM}$ distribution (bottom left), and the thrust ($\tau = 1-T$) event-shape distribution, in comparison to the same reference measurements~\cite{Achard:2004sv,Tanabashi:2018oca} that were used for the Monash tune. The green and yellow shading in the ratio panes denote $1\sigma$ and $2\sigma$ experimental uncertainties respectively, with systematic and statistical uncertainties added in quadrature. (Inside the green band, the statistical component by itself is indicated by a slightly lighter green shading.)

\begin{figure}[tp]
\centering
\includegraphics[width=0.6\textwidth]{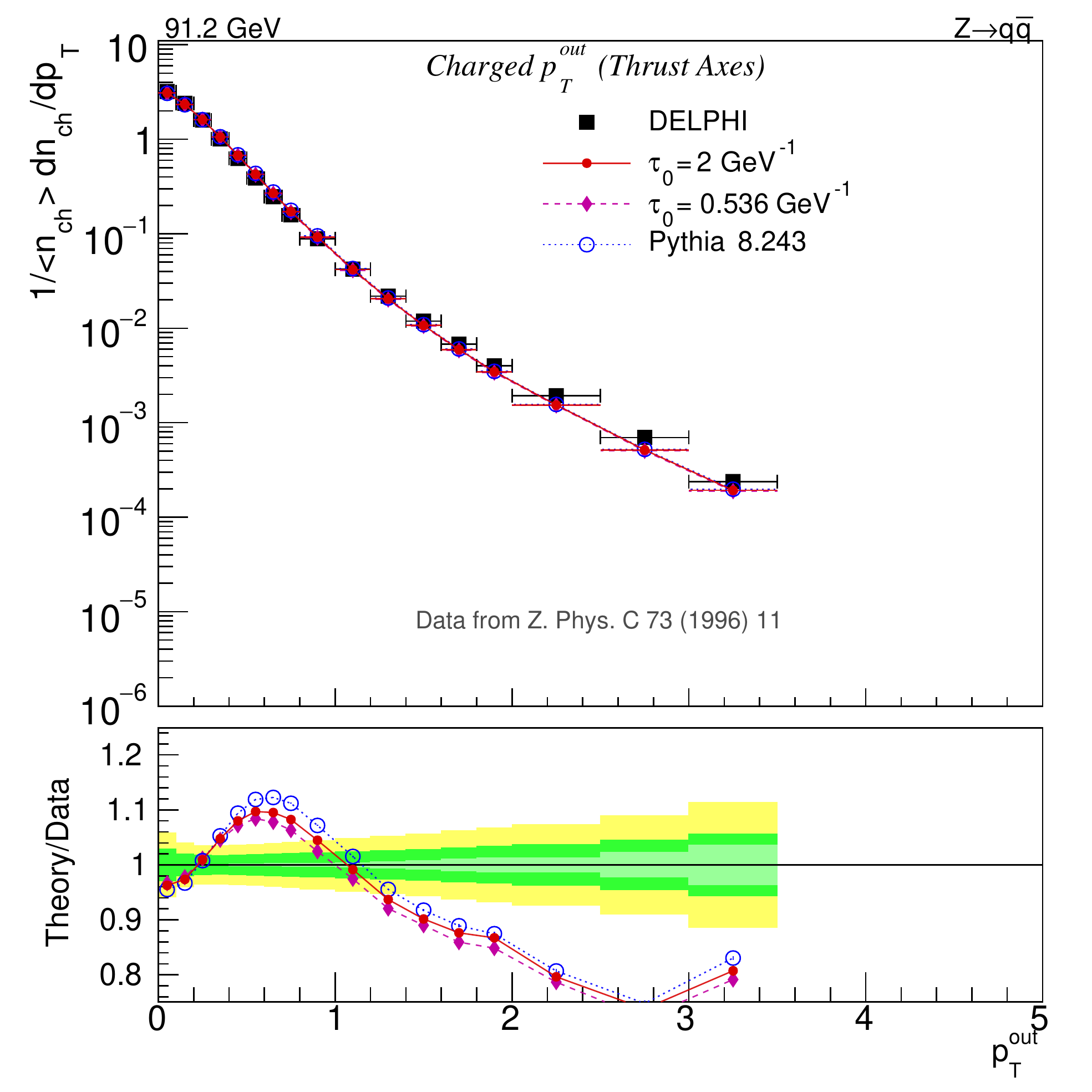}
\caption{The out-of-plane charged-particle $p_\perp$ distribution, as measured along the Thrust minor axis. Data from~\cite{Abreu:1996na}.
\label{fig:pTout}}
\end{figure}

By contrast, a measured distribution which does exhibit some change as a result of our modelling is the out-of-plane $p_\perp$ distribution (with similar but smaller changes for the in-plane ditto), shown in fig.~\ref{fig:pTout}, compared to a DELPHI measurement~\cite{Abreu:1996na}. (Specifically, the plotted quantity is the charged-hadron momentum component along the Thrust minor axis.)
We emphasise that the $p_{\perp\mathrm{out}}$ distribution is among the most challenging observables for all event generators\footnote{See, e.g., the comparisons available on the \url{mcplots.cern.ch} web site~\cite{Karneyeu:2013aha}).}. In the hard tail beyond $p_{\perp\mathrm{out}} \sim 1\,\mathrm{GeV}$ this probably involves an interplay with perturbative physics beyond ${\cal O}(\alpha_s^2)$; more interesting from the perspective of non-perturbative modelling is the peak of the spectrum in the region below 1 GeV. In this region, PYTHIA's default modelling (blue open circles) results in a slightly too hard spectrum, with too many hadrons in the range $p_{\perp\mathrm{out}}\in[0.4,1]$ GeV and not quite enough at smaller $p_{\perp\mathrm{out}}$ values. This feature is closely linked to the width of the Gaussian $p_\perp$ distribution for string breaks. 

While our model does not change the hard tail of the distribution beyond 1 GeV, the non-perturbative component does change, becoming broader, with  hadrons associated with very early string breaks acquiring higher average $p_\perp$ values, while others, associated with late string breaks, become somewhat softer. (At the very lowest $p_{\perp\mathrm{out}}$ values, however, it remains dominated by $p_\perp$ kicks generated by hadron decays.) 

If anything, the changes produced by the two toy scenarios we include here are too mild, possibly motivating investigations of more dramatic shape changes. Any firm conclusion on this would, however, require addressing the deficit in the hard tail as well, which is beyond the scope of our study. An alternative, which we elaborate on a bit further below, could be to separate the hard and soft components of the measured distribution by excluding events with more than 2 jets and/or measuring $p_\perp$ with respect to a local axis for each jet. 

\begin{figure}[tp]
    \centering
\begin{tabular}{cc}%
\hspace*{-3.5mm}\includegraphics[width=0.5\textwidth]{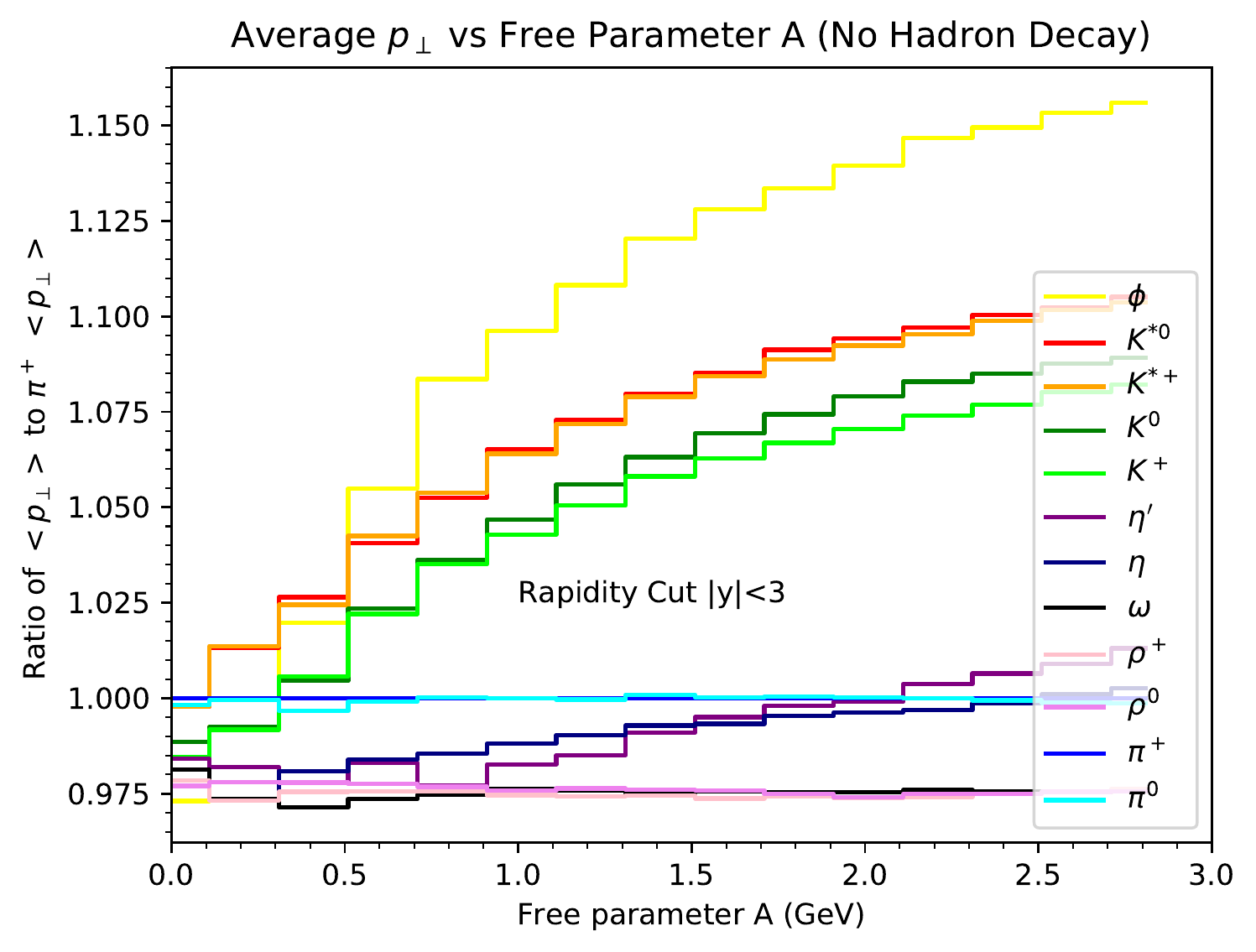}\hspace*{-1mm}&\hspace*{-1mm}%
\includegraphics[width=0.5\textwidth]{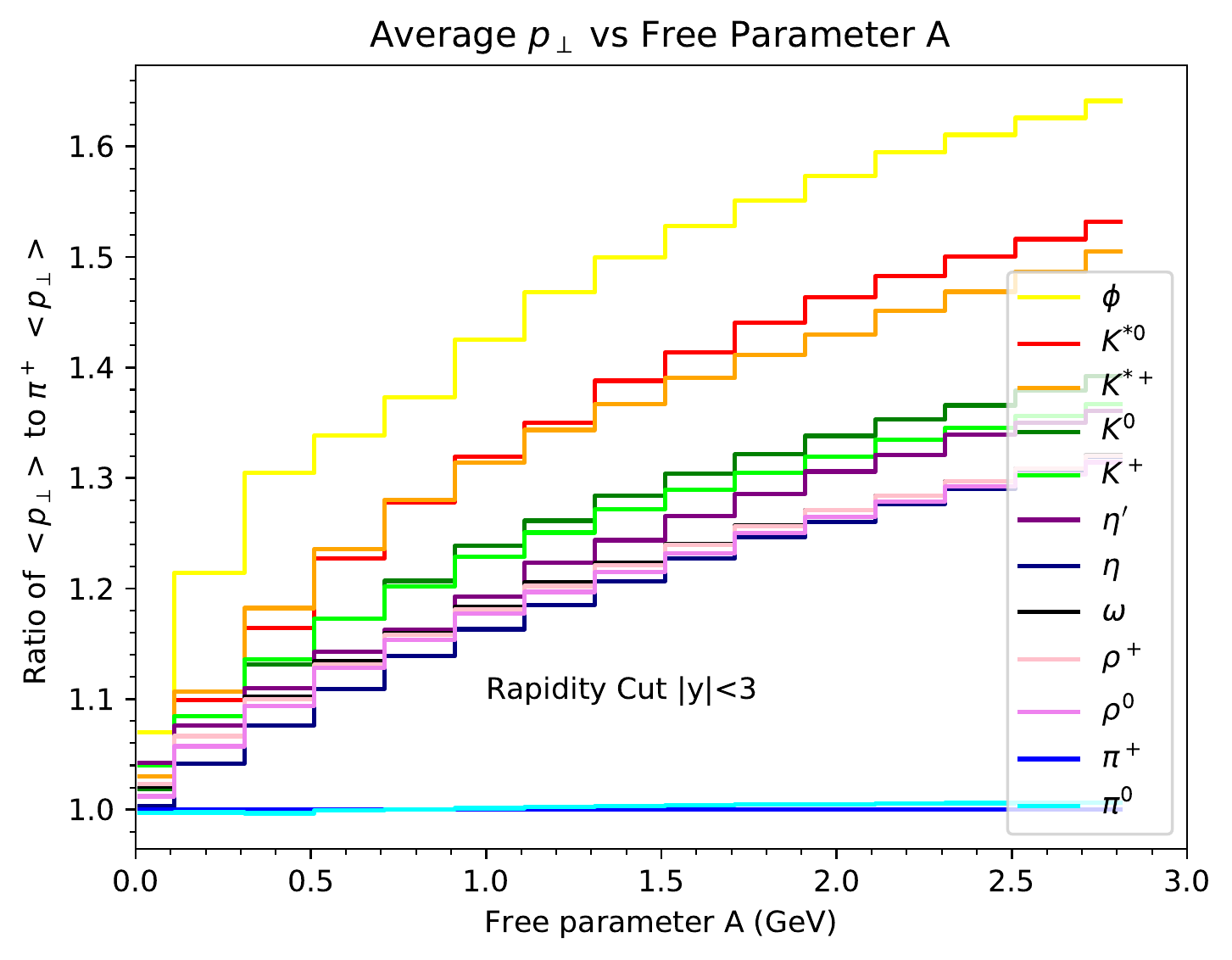}\hspace*{-3mm}%
\end{tabular}
        \caption{Ratio $\left<p_\perp\right>_\mathrm{Meson} / \left<p_\perp\right>_{\pi^+}$ for various meson species, as a function of $A$ for fixed $\tau_0 = 2$ GeV$^{-1}$, before (left) and after (right) including the effects of hadron decays. In both plots, a cut requiring $|y|<3$ was imposed to suppress endpoint effects.}
    \label{fig:pTvsA}
\end{figure}

More information can be gained, however, if we incorporate flavour and multiplicity dependence of the $p_\perp$ distributions, instead of just looking at charged particles in general. While the baseline Schwinger model predicts that all string breaks have the same (Gaussian) $p_\perp$ spectrum, regardless of flavour, a generic aspect of many alternative scenarios, including ours, is that heavier quarks (i.e., strange ones) should be associated with higher $p_\perp$ values. 
This is illustrated by the plots in fig.~\ref{fig:pTvsA}, which show the average $p_\perp$ of various meson types, relative to the average $p_\perp$ of pions, as a function of the free model parameter, $A$, in the range $[0,3]\,\mathrm{GeV}$, for fixed $\tau_0=2\,\mathrm{GeV}^{-1}$. 
The left-hand edge of the plots corresponds to the baseline (Monash 2013) modelling, while the right-hand edge ($A=3\,\mathrm{GeV}$) represents an extreme scenario not really compatible with non-perturbative physics but included to illustrate the asymptotics of the model. (We expect realistic $A$ values to be  below 1 GeV.) 
A rapidity cut requiring $|y|<3$ (with respect to the string axis) was imposed to suppress the influence of the first-rank hadrons which (due to the absence of a non-perturbative $p_\perp$ kick to the endpoint quarks and the flavour dependence of the $Z^0\to q\bar{q}$ branching fractions) exhibit 
differences between otherwise nearly related particle types such as $K^{*0}$ and $K^{*\pm}$.

Focusing first on the left-hand plot, which shows the  distributions for primary hadrons, i.e., without including the effects of hadron decays, we observe that the $p_\perp$ of primary strange mesons all increase, relative to the pions, while those of non-strange ones are approximately independent of $A$. Thus, even if the model were retuned (as will be done below) to preserve the reference value for, e.g., the average charged-particle $p_\perp$, our model (and any others of a similar ilk) will still predict a different pattern of dependence on hadron mass and strangeness; it should be possible to determine this pattern experimentally in archival $ee$ data.

A further feature that can be noted in the left-hand plot is a slight tapering off towards the largest $A$ values. This is due to phase-space suppression. We are here running at $\sqrt{s} = m_Z$, and as $A$ becomes very large the phase space starts to be slightly constraining. (This conclusion was verified by noting that the tapering goes away for very large $\sqrt{s}$.) 

The right-hand plot shows what happens when we include the effect of hadron decays. This smears out the distributions, moreover we see the  $\left<p_\perp\right>$ of non-strange hadrons (specifically, $\rho^0$,$\rho^\pm$ and $\omega$) beginning to scale  with $A$ as well,  relative to pions. This is mainly due to the fact that about half of pions come from hadron decays, hence their $\left<p_\perp\right>$ increases much less quickly with $A$ in absolute terms, steepening all other curves relative to the pions. 

\begin{figure}[tp]
    \centering
    \includegraphics[width=0.6\textwidth]{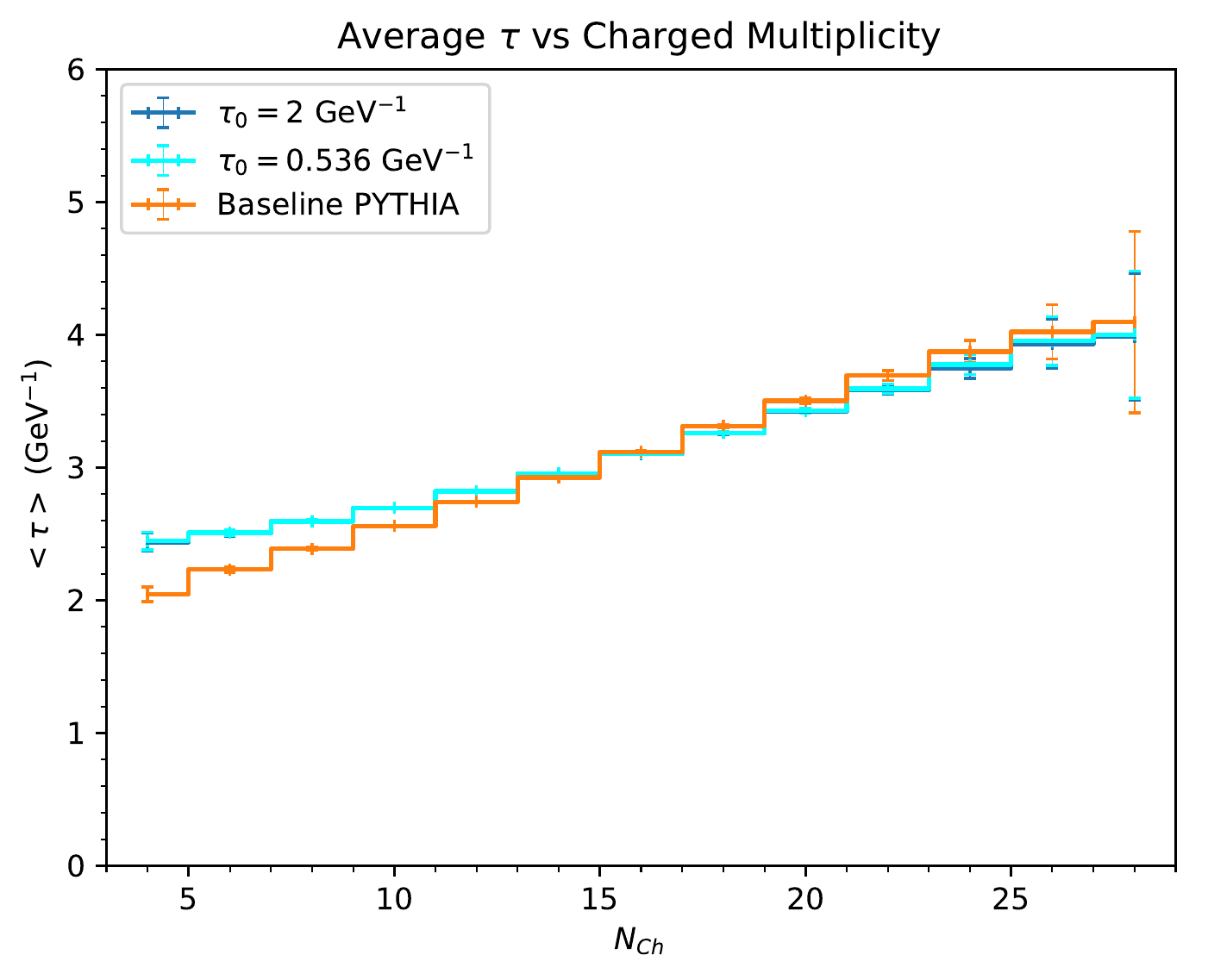}
    \caption{Variation in average $\tau$ with charged multiplicity for our model (blue) and baseline PYTHIA (orange).}
    \label{fig:TauVsMultiplicity}
\end{figure}

The hallmark feature of our model is of course the postulate that string breakup properties could have a $\tau$ dependence. In the absence of techniques for femtometre-scale vertexing, however, the actual $\tau$ values for string breaks are not measurable quantities. The next-best thing is a physical observable that is correlated with $\tau$. One such,  suggested to us in discussions with T.~Sj\"ostrand, is plotted in fig.~\ref{fig:TauVsMultiplicity}: the average multiplicity of charged hadrons. In both the baseline model and in our $\tau$-dependent variants there is clearly an approximately linear relationship between $\left<\tau\right>$ and $\left<N_\mathrm{Ch}\right>$. The physics behind this is that a string break at a low $\tau$ value 
removes a relatively large fraction of the forward light cone that would otherwise have been available for further string breaks; thus systems with low-$\tau$ breaks should have a  lower-than-average total number of string breaks (and hence smaller numbers of hadrons). Within the general paradigm of the string model, therefore, for a given string invariant mass, low-$N_\mathrm{Ch}$ events will exhibit an enhancement of low-$\tau$ breakups relative to high-$N_\mathrm{Ch}$ ones.

While the $N_\mathrm{Ch}$ dependence of $pp$ phenomenology has been extensively investigated, it is completely different from that of $e^+e^-$ annihilations since the main drivers for multiplicity fluctuations in $pp$ collisions are the numbers and invariant masses of strings produced via multi-parton interactions, not the fluctuations of individual strings of fixed invariant mass; see e.g.~\cite{Sjostrand:1987su}. 
To our knowledge, the $N_\mathrm{Ch}$ dependence of $ee$ collision properties has not been similarly investigated. We are therefore not in a position to compare to measurements, but instead highlight the opportunities we see, and use our model to illustrate them.

\begin{figure}[tp]
    \centering
    \includegraphics[height=\textwidth,width=0.95\textwidth]{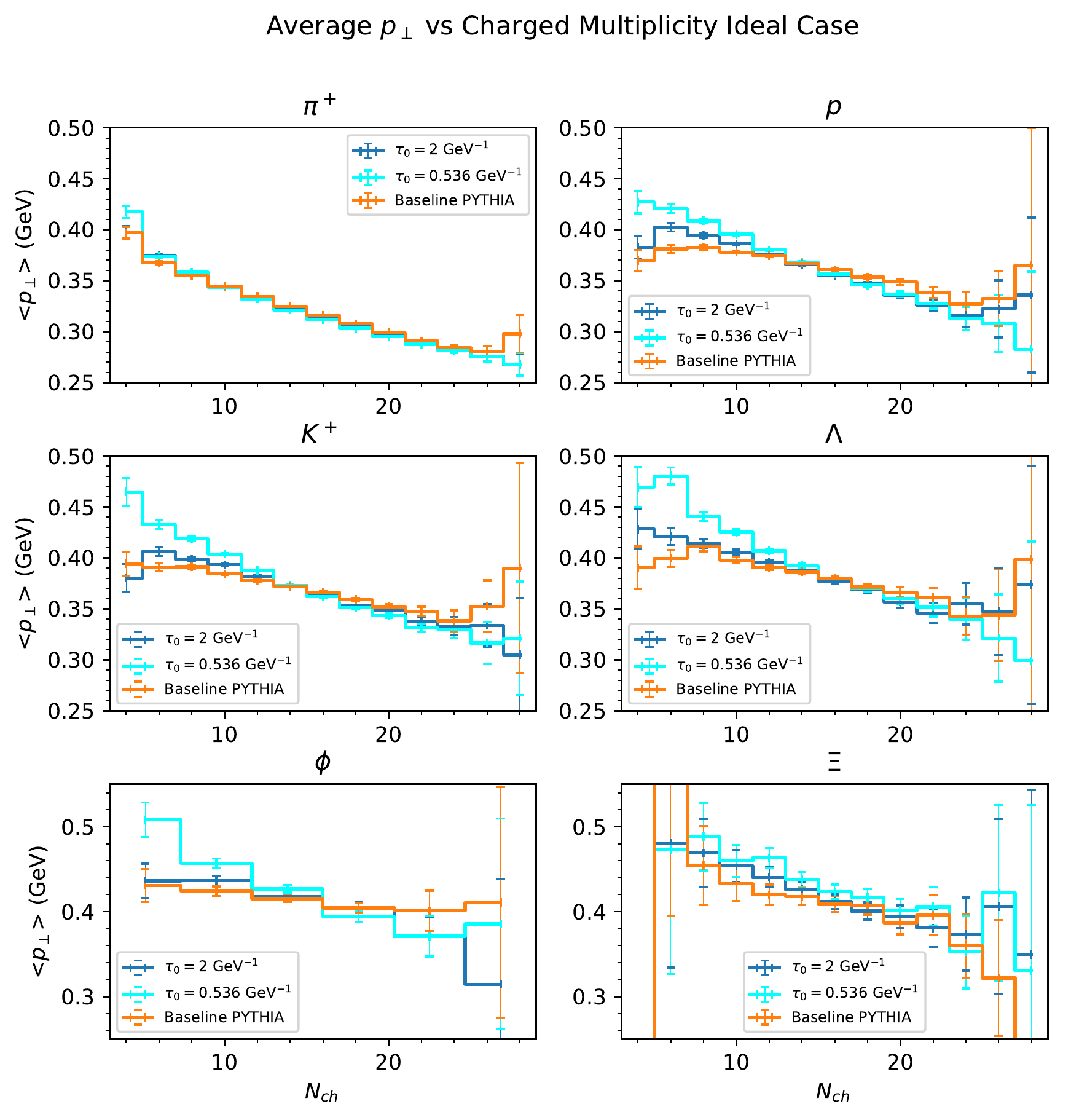}
    \caption{Mean $p_\perp$ versus charged multiplicity for $\pi^+$, $p$, $K^+$, $\Lambda$, $\phi$ and $\Xi$.} 
    \label{fig:pTvsMultIdeal}
\end{figure}

Within the context of our model, low $\tau$ implies a higher string tension and hence higher $p_\perp$ and strangeness fractions. We would therefore expect these features to be enhanced in low-multiplicity $e^+e^- \to \mathrm{hadrons}$ events, relative to the baseline model. As a first step towards testing this, we study an idealised case in which the string axis is forced to lie upon the $z$ axis, so that there is no ambiguity in which axis to choose. Comparisons between our model and baseline PYTHIA were performed for charged pions (no strange content), charged kaons (1 strange quark) and $\phi$ mesons (2 strange quarks). In the baryon sector, the same comparisons were generated for protons (no strange content), as well as $\Lambda$ (1 strange quark), charged $\Sigma$ (1 strange quark) and charged $\Xi$ (2 strange quarks), yielding the plot shown in fig.~\ref{fig:pTvsMultIdeal}.

The first obvious trend is that all the models, irrespective of $\tau$ dependence, exhibit decreasing tendencies of $\left<p_\perp\right>$ with $N_\mathrm{Ch}$, the opposite of what is observed in $pp$ collisions. This is a consequence of energy conservation; in $e^+e^- \to \mathrm{hadrons}$, the total invariant mass is fixed (the $Z$ mass) and hence if we want to produce \emph{many} hadrons, each of them must have somewhat smaller momenta. 

Looking beyond this overall trend, differences between the models emerge, especially at low $N_\mathrm{Ch}$ for strange hadrons. While the $\tau_0 = 2$ GeV$^{-1}$ variant does not have much influence on the $\left<p_\perp\right>$ vs multiplicity spectra, 
the $\tau_0 = 0.536$ GeV$^{-1}$ variant which allows a large increase in the effective string tension at very early times, predicts a statistically significant effect, with the $\phi$ mesons exhibiting the steepest gradient compared to the relatively flat prediction of base PYTHIA. Unfortunately the LEP-sized samples used for this study appear to contain too few $\Xi$ baryons (cf.~the lower right-hand pane of fig.~\ref{fig:pTvsMultIdeal}) for these to furnish a useful additional test of the scaling with the number of strange quarks, at least in the context of the effects produced by the scenarios considered here. 

\begin{figure}[tp]
    \centering
    \includegraphics[height=\textwidth,width=0.95\textwidth]{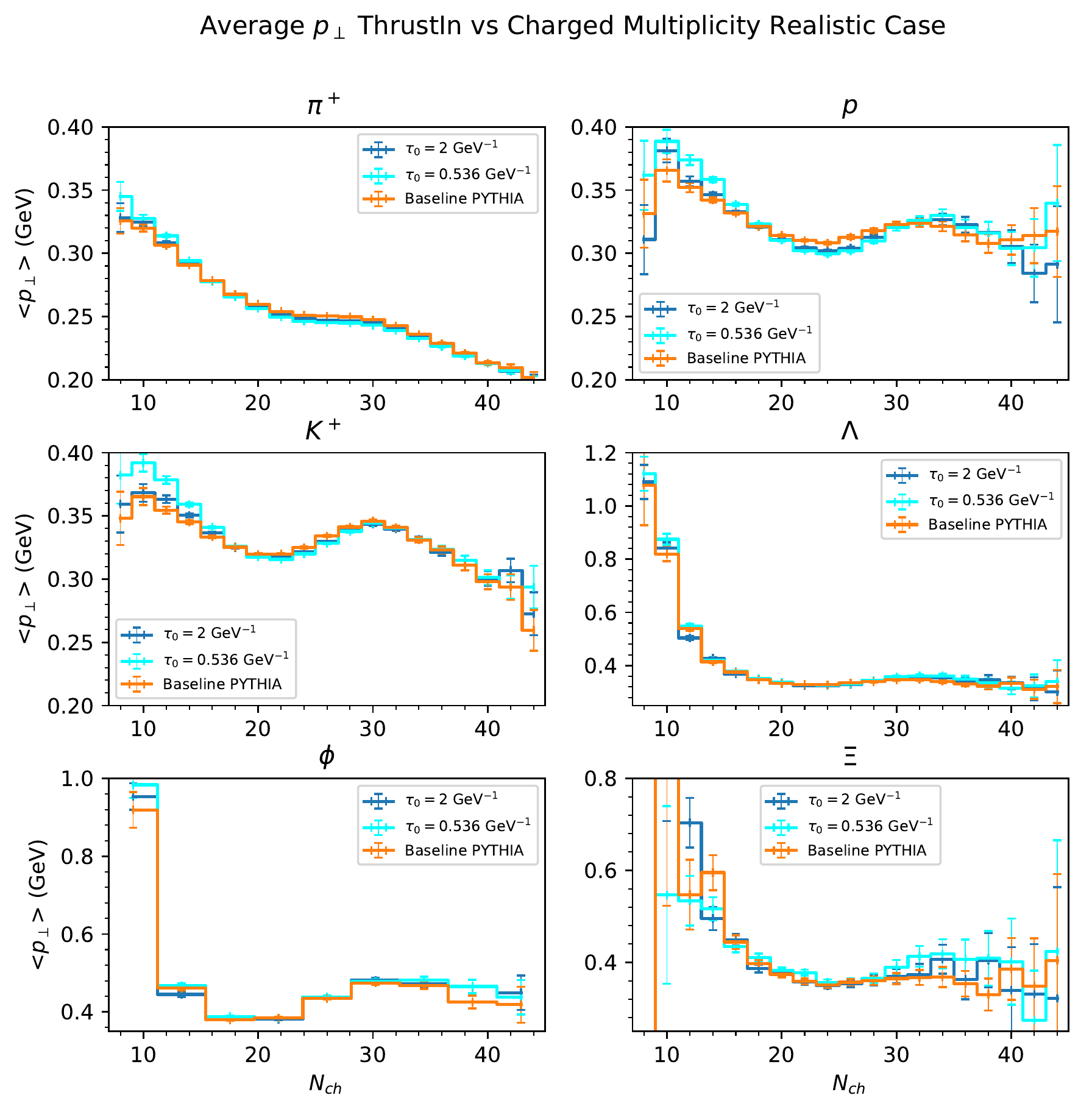}
    \caption{Mean $p_\perp$ along thrust major axis versus charged multiplicity for $\pi^+$, $p$, $K^+$, $\Lambda$, $\phi$ and $\Xi$.}
    \label{fig:pTvsMultReal}
\end{figure}

Turning now to a more realistic case in which the string axis is a priori unknown, we measure the $p_\perp$ with respect to the Thrust axis, and show the in-plane $p_\perp$ component in fig.~\ref{fig:pTvsMultReal}. 

It is evident that the ambiguity in the choice of axis somewhat smears the picture, but again especially the Kaon spectrum, and to some extent the $\phi$ and $\Lambda$ ones, appear to offer discriminatory power. We expect that an experimental analysis could be optimised beyond what we are doing here by systematically comparing different options for how to choose the axis (e.g., thrust vs sphericity vs jet axes), measuring $p_\perp$ values with respect to individual jet axes or excluding 3-jet events to make the global axis choice more appropriate, focusing on mid-rapidity hadrons that are unlikely to be contaminated by the original endpoint quarks from the $Z\to q\bar{q}$ decay, and/or even trying to isolate the two hadrons stemming from a specific low-$\tau$ break by looking at relative $p_\perp$ differences between neighbouring pairs of hadrons in rapidity space. 

We note that the discriminatory power highlighted in the previous two figures diminishes if one uses the out-of-plane $p_\perp$ component instead of the in-plane one. A possible reason for this is that low-multiplicity events are already dominated by 2-jet events in which the perturbative activity barely defines a plane, so that the non-perturbative corrections can be significant in determining what is ``in'' and what is ``out''. If so, a single large $p_\perp$ value generated by a non-perturbative breakup would show up in  $\left<p_{\perp\mathrm{in}}\right>$ but not in $\left<p_{\perp\mathrm{out}}\right>$. 

\begin{figure}[tp]
    \centering
    \includegraphics[height=0.9\textwidth,width=0.95\textwidth]{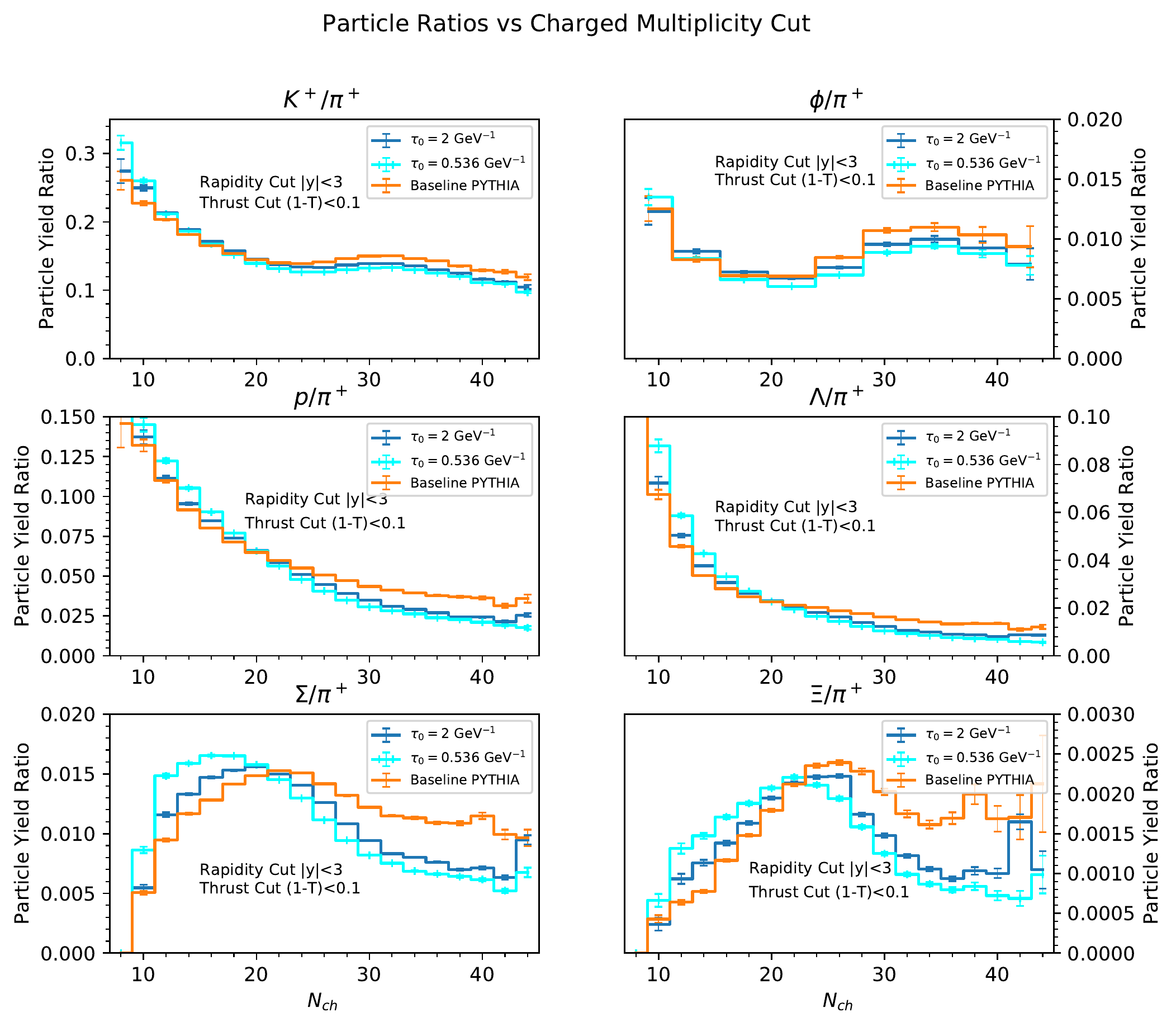}
    \caption{Particle yields as a ratio to pions for $K^+$, $\phi$, $p$, $\Lambda$, $\Sigma$ and $\Xi$ after cuts.}
    \label{fig:RatioCut}
\end{figure}

As our final examples of salient distributions that could be measured in archival $ee$ data, we show the hadron$/\pi$ distributions for different hadron species as functions of $N_\mathrm{Ch}$ in fig.~\ref{fig:RatioCut}. To suppress effects of the original $Z\to q\bar{q}$ endpoint quarks, we include only particles with rapidities $|y|<3$ with respect to the Thrust axis, for events with low values of $1-T \le 0.1$ , i.e., reasonably pencil-like events for which the Thrust axis should provide a fairly good global axis choice. The number of particles remaining after both of these cuts is reduced by around 36\%. The relationships between particle yield ratio and charged multiplicity for these hadrons are shown in fig.~\ref{fig:RatioCut}.

At low multiplicities, we see higher strangeness fractions, reflecting the earlier $\left<\tau\right>$ values. This trend is particularly pronounced for strange baryons such as $\Sigma$ and $\Xi$ shown in the bottom two panes. This plot indicates that effects such as those represented in our model can have a significant effect on the correlation between strangeness and particle multiplicity. Generically, if earlier times are associated with higher scales, our prediction is for \emph{higher} average $p_\perp$ and strangeness fractions at lower multiplicities, the opposite of the trend observed for $pp$ collisions. However, as already mentioned the overall main driving factor for the behaviour in $ee$ is the fixed total invariant mass, which does not carry over directly to $pp$. A dedicated study of $pp$ phenomenology is, however, beyond the scope of this work\footnote{This is partly due to a purely technical limitation to do with how PYTHIA's \texttt{UserHooks} framework is currently structured, which causes our model to run about an order of magnitude slower than the standard fragmentation model.}.
The main point we would wish to make in connection with $pp$ phenomenology is the following: that the modelling of individual strings (as in $ee$) is the starting point for the modelling of several ones, and hence our ability to draw firm \emph{quantitative} conclusions from the $pp$ data will ultimately depend on how thoroughly we think we 
understand the reference case of a single isolated string.

\section{Conclusions and Outlook}
\label{sec:conclusions}

In this work, we have developed a toy-model extension to the conventional Lund string fragmentation model, in which we allow for the string tension, $\kappa$, to depend on proper time. The main inspiration comes from recent arguments~\cite{Berges:2017hne} that an expanding string may exhibit thermal excitations with a characteristic temperature that depends inversely on proper time. We also consider a variant inspired by the Coulomb term in the Cornell potential.

Such effects would lead to correlations between the proper times of string breakups and their $p_\perp$ broadening and strangeness fractions, which are absent from the baseline Lund model. While the proper time is not directly measurable, we demonstrate that the average charged multiplicity can be used as a proxy, at least in the context of a fixed total invariant mass of the string, such as in $e^+e^- \to \mathrm{hadrons}$. 

We are not aware of an experimental study in $ee$ collisions of the dependence of average $p_\perp$ or strangeness on charged multiplicity as we have studied here. We believe that such an analysis would be worthwhile to constrain not only the scenarios considered here but any model of a similar ilk. 

Given the correlation we have observed between strange particle production and charged multiplicity, our model could be extended to $pp$ collisions, where a clear dependence between these two quantities has been observed at the LHC ~\cite{ALICE:2017jyt}. The claim is not that the effects incorporated in our model could account for the LHC observations, if anything the effect we predict in the $ee$ context is of opposite sign compared with that seen in the LHC data, but that the we think the (uncertainties on the) properties of the single string in isolation may be relevant to our interpretation of the collective effects seen in $pp$.

An interesting extension to our model could be to allow for background fields to modify the effective tension as well; this could provide an avenue for incorporating collective effects in this framework, possibly in conjunction with asymmetric contributions to the $p_\perp$ distributions to represent repulsion effects along a similar vein as recently proposed in~\cite{Duncan:2019poz}. 

\subsection*{Acknowledgements}
We are indebted to G.~Gustafson for valuable discussions and comments on this work. 
This work was funded in part by the Australian Research Council via Discovery Project DP170100708 -- “Emergent Phenomena in Quantum Chromodynamics”. This work was also supported in part by the European Union's Horizon 2020 research and innovation programme under the Marie Sklodowska-Curie grant agreement No 722104 -- MCnetITN3.

\begin{appendices}
    \section{Constrained values of $A$ and $\tau_0$} \label{appendix:Avstau0}
With the $\tau$ distribution obtained for $ee\to \mathrm{hadrons}$ using default settings of PYTHIA, see fig.~\ref{fig:kappaTau}, the constraint $\left<\kappa(\tau)\right> = \kappa_\mathrm{tune} = 0.353\,\mathrm{GeV}^2$ implies the following relationship between $\tau_0$ and $A$. 
    \begin{table}[h!]
        \centering
        \begin{tabular}{|c|c|}
            \hline
            $\tau_0$ (GeV$^{-1}$) & $A$ (GeV)\\
            \hline \hline
            0.1 & 0.381\\
            \hline
            0.2 & 0.382\\
            \hline
            0.3 & 0.383\\
            \hline
            0.4 & 0.386\\
            \hline
            0.5 & 0.389\\
            \hline
            0.6 & 0.393\\
            \hline
            0.7 & 0.397\\
            \hline
            0.8 & 0.401\\
            \hline
            0.9 & 0.406\\
            \hline
            1 & 0.412\\
            \hline
            1.1 & 0.418\\
            \hline
            1.2 & 0.424\\
            \hline
            1.3 & 0.430\\
            \hline
            1.4 & 0.437\\
            \hline
            1.5 & 0.444\\
            \hline
            1.6 & 0.452\\
            \hline
            1.7 & 0.459\\
            \hline
            1.8 & 0.467\\
            \hline
            1.9 & 0.476\\
            \hline
            2 & 0.484\\
            \hline
            2.1 & 0.493\\
            \hline
            2.2 & 0.503\\
            \hline
            2.3 & 0.512\\
            \hline
            2.4 & 0.522\\
            \hline
            2.5 & 0.533\\
            \hline
        \end{tabular}
        \caption{Constrained $A$ value for several values of $\tau_0$ subject to $<\kappa(\tau)> = \kappa_{Tune}$.}
        \label{tab:Avstau0}
    \end{table}
\end{appendices}
\clearpage
\bibliography{main}

\end{document}